\documentclass[a4paper,11pt,english]{article}
\usepackage[latin1]{inputenc}
\usepackage[english]{babel}
\usepackage{graphicx}

\usepackage{verbatim,times,bbm}
\usepackage{color}

\usepackage{amsmath,amssymb}

\usepackage[bindingoffset=0.5cm,left=2.5cm,top=3cm,bottom=3cm,right=2.5cm]{geometry} %,nohead,nofoot]{geometry}

\begin{document}

\title{Robustness of geometric phase under parametric noise}

\author{Cosmo Lupo\\
Dipartimento di Fisica, Universit\`a di Camerino, I-62032 Camerino,
Italy \\
{\normalsize cosmo.lupo@unicam.it}
\and Paolo Aniello \\
Dipartimento di Scienze Fisiche dell'Universit\`a di Napoli ``Federico II'' \\
\& INFN -- Sezione di Napoli \\
\& Facolt\`a di Scienze Biotecnologiche Universit\`a di Napoli ``Federico II'', \\
via Cintia I-80126 Napoli, Italy}

\maketitle

\begin{abstract}
We study the robustness of geometric phase in the presence of
parametric noise. For that purpose we consider a simple case study,
namely a semiclassical particle which moves adiabatically along a
closed loop in a static magnetic field acquiring the Dirac phase.
Parametric noise comes from the interaction with a classical
environment which adds a Brownian component to the path followed by
the particle. After defining a gauge invariant Dirac phase, we
discuss the first and second moments of the distribution of the
Dirac phase angle coming from the noisy trajectory.

\vspace{0.5cm} PACS: 03.65.Vf, 02.50.Ey, 03.67.-a
\end{abstract}

% 03.65.Vf    Phases: geometric; dynamic or topological
% 02.50.Ey    Stochastic processes
% 03.67.-a    Quantum information

\section{Introduction}

The first reference to the role played by geometric phases in
physics dates back to the work of S.\ Pancharatnam \cite{Panch} in
the context of interferometry of polarized beams of light. Later the
same phenomenon was described by M.\ V.\ Berry \cite{Berry} for
quantum mechanical systems in the adiabatic limit. A mathematical
insight into its origin was provided by B.\ Simon \cite{Simon} which
recognized that Berry phases could be interpreted as holonomies on a
fiber bundle. Subsequently, quantum geometric phases (i.e.\ {\it
quantum holonomies}) have been predicted and observed in various
physical systems, and several generalizations and extensions were
proposed \cite{Wilczek,Aharonov,Samuel} (see \cite{Bohm} and the
references therein). Generally speaking, we may say that geometric
phases appear in correspondence with a cyclic evolution of a
relevant Hilbert (sub)space. The dimension of the cyclic (sub)space
determines the features of the corresponding geometric phase:
Abelian, i.e.\ $\mathrm{U}(1)$-valued, for Hilbert space of unit
dimension, nonAbelian, i.e.\ $\mathrm{U}(N)$-valued, in the case of
$N$-dimensional cyclic Hilbert space.

Few years after the seminal papers by M.\ V.\ Berry and B.\ Simon,
the scientific community discovered the potentialities of quantum
mechanical systems in the context of information and communication
technology \cite{Bennett}, leading to the birth of {\it quantum
information science} \cite{Nielsen_Chuang}. Since information
processing obeys physical laws, a reversible quantum algorithm is
represented by a unitary transformation as long as information is
encoded as vectors in a Hilbert space. It follows that quantum
holonomies, being unitary transformations, can serve as quantum
logical gates to implement quantum algorithms. This idea was first
proposed and discussed in \cite{Zanardi_Rasetti}, where the authors
also proved that universal computation \cite{Deutsch} can be in
general realized by means of solely (nonAbelian) geometric phases. A
vast literature followed, which included both theoretical proposals
\cite{Falci,Duan} and experimental realizations \cite{nist,montana}
of simple {\it holonomic quantum gates}.

In view of the application for quantum information processing, one
usually considers the case of adiabatic geometric phase (also called
Berry phase). In this case the system Hamiltonian is supposed to be
a smooth function of the coordinates on a suitable manifold, often
called the `parameter manifold'. The geometric phase arises in the
adiabatic approximation in correspondence with a closed loop in the
parameter manifold.

It is worth remarking, however, that that geometric approach to the
computation can be rather demanding from a technological point of
view. Nevertheless, its advantage with respect to the standard
dynamical approaches relies on the fact that Berry phases are argued
to be particular robust with respect to noise. In particular, it is
argued that holonomic quantum gates can be robust with respect to
certain kind of {\it classical parametric noise} \cite{Jones}. This
kind of noise can be modeled as coming from the interaction with a
classical environment.

The robustness of quantum logic gates is indeed a crucial issue
because of the inherent fragility of quantum mechanical systems.
Hence, although quantum error correction protocols exist \cite{ECC},
quantum gates which are {\it a priori} robust are welcome. It is
worthwhile to mention another remarkable proposal for a fault
tolerant computation, namely {\it topological} computation
\cite{Kitaev}. This approach is based on nonAbelian Aharonov-Bohm
topological phases.

As we have anticipated, the geometric phase is believed to be robust
with respect to classical parametric noise. This argument has been
the central issue of several investigations, in particular we
mention the work by the G.\ De Chiara and M.\ G.\ Palma
\cite{DeChiara}, as well as the results presented in \cite{Solinas}.
Several physical models have been taken into consideration to the
study of the robustness of geometric phase. Recently, the robustness
of geometric phase has been tested experimentally in trapped
polarized ultra-cold neutrons \cite{Filipp}. However it is worth
noticing that this issue is largely independent of the details of
the physical model under consideration. For this reason, our
discussion will be devoted to the simplest settings in which
geometric phases appear. In the following sections we consider a
charged semiclassical particle which is adiabatically and cyclically
moved in a static magnetic field. The particle acquires the {\it
Dirac phase} which is proportional to the magnetic flux enclosed by
the particle trajectory. Two different settings will be described:
in the first case the semiclassical particle is in the presence of a
homogeneous magnetic field; in the second example the particle is
subject to the field generated by a magnetic monopole.

\section{Dirac phase under noise}

In the following sections we consider the effects of parametric
noise on quantum holonomies in two simple but remarkable examples.
We consider as well two models for the noise, respectively
represented by a Wiener and by a Ornstein-Uhlenbeck process. The
examples involve a simple physical system made of a charged
semiclassical particle in the presence of a static magnetic field
$\mathbf{B}\equiv(B_x,B_y,B_z)$. We indicate the corresponding
vector potential as $\mathbf{A}\equiv(A_x,A_y,A_z)$ and the particle
position as $\mathbf{r}\equiv(x,y,z)$.

Let us initially consider the case in which the particle follows a
closed loop during a certain operational time $T$, hence we have
$\mathbf{r}_0(T)=\mathbf{r}_0(0)$. In the adiabatic approximation,
if the internal state of the particle in initially described by the
a state vector $\Psi(0)$, after the closed loop in the parameter
space the internal state of the particle will be described by the
vector $\Psi(T)=e^{-i\phi_g}\Psi(0)$. Where
\begin{equation}\label{flux}
\phi_g = - \frac{q}{\hbar} \oint_{\mathbf{r}_0} \mathbf{A} \cdot
d\mathbf{r}
\end{equation}
is the acquired Dirac phase. Here we neglect the dynamical
contribution to the phase factor. Let now the particle be subjected
to Brownian motion due to the interaction with a classical
environment. In this case a stochastic component
$\mathbf{r}_\mathrm{n}(t)$ adds to the drift motion leading to the
noisy trajectory
\begin{equation}\label{noisypath}
\mathbf{r}(t) = \mathbf{r}_0(t) + \mathbf{r}_\mathrm{n}(t).
\end{equation}
Our aim is to study the Dirac phase acquired as a consequence of the
noisy trajectory. First of all, we need a well defined notion of
Dirac phase for the Brownian trajectory. The point is that the noisy
path is in general noncyclic, while gauge invariance of the phase
(\ref{flux}) requires a closed loop. As we discuss below, it is
possible to define a gauge invariant Dirac phase in a rather natural
way. Secondly, since the Dirac phase is determined by the stochastic
trajectory of the particle, we expect that the resulting phase is a
stochastic variable as well. Our interest will be focalized on its
mean value and variance.

To conclude this section, we introduce our definition of Dirac phase
for noncyclic trajectories. From a general point of view, it is
possible to define the geometric phase in correspondence of
noncyclic evolution if one introduces a rule that allows to close an
open loop. As it was discussed in \cite{Samuel}, the natural choice
is to close the loop with a geodesic curve, where the metric is
defined by the hermitian product in the Hilbert space which is
relevant in the given context. In our case, the natural choice is to
consider the Euclidean metric in $\mathbb{R}^3$, hence we consider
the following definition of gauge invariant Dirac phase:
$$
\phi_g = \int_\mathbf{r} \mathbf{A}\cdot d\mathbf{r} + \int_G
\mathbf{A}\cdot d\mathbf{r},
$$
where $G$ indicates the straight line joining the final point
$\mathbf{r}(T)$ to the initial point $\mathbf{r}(0)$.

\section{Particle in homogenous magnetic field}

In this section we consider the case of a static homogeneous
magnetic field $\mathbf{B}\equiv(0,0,B)$. We can write the vector
potential in the asymmetric gauge as $\mathbf{A}\equiv(-yB,0,0)$,
hence the Dirac phase is determined by the integral
\begin{equation}\label{nongauge}
\phi_g = \int_0^T y(t) dx(t).
\end{equation}
We are going to consider a noisy path of the form (\ref{noisypath})
with a trivial drift component
$$
\mathbf{r}_0(t)\equiv(x(0),y(0),z(0))
$$
yielding a trivial noiseless
Dirac phase. For the sake of simplicity we consider the motion of
the semiclassical particle as confined in the plane $x-y$,
perpendicular to the magnetic field.

\subsection{Wiener process}

As a first model for the noise component, we consider the case of a
Wiener process. Hence we impose the following conditions on
two-times correlation functions of the Brownian component
$\mathbf{r}_\mathrm{n}(t)\equiv(x(t),y(t))$:
\begin{eqnarray}
\langle x(s) x(t) \rangle & = & B_x \delta(s-t), \label{xx}\\
\langle y(s) y(t) \rangle & = & B_y \delta(s-t), \label{yy}\\
\langle x(s) y(t) \rangle & = & 0 \label{xy}.
\end{eqnarray}
The noisy path has initial point $x(0)$, $y(0)$ and final point
$x(T)$, $y(T)$. In order to ensure gauge invariance, we add one term
to the equation (\ref{nongauge}), obtaining the following expression
for the gauge invariant Dirac phase angle:
\begin{equation}\label{gauge}
\phi_g = \int_0^T y(t) dx(t) + \frac{1}{2}\left[ x(0) - x(T) \right]
\left[ y(0) + y(T) \right] = \int_0^T y(t) dx(t) +
\frac{1}{2}\Delta\Sigma,
\end{equation}
where $\Delta := x(0) - x(T)$ and $\Sigma := y(0) + y(T)$. With this
definition we can compute the corresponding mean value and variance
of the phase factor $e^{i\phi_g}$. However, for small perturbation
we can consider directly the mean and variance of the angle
$\phi_g$. Concerning the mean value, it is immediate to see that it
vanishes as long as the two processes $x(t)$ and $y(t)$ are
statistically independent. On the other hand, the variance reads
$$
\sigma_g^2 = \langle \int_0^T y(t)dx(t) \int_0^T y(s) dx(s) \rangle
+ \langle \Sigma \Delta \int_0^T y(t)dx(t) \rangle + \frac{1}{4}
\langle \Sigma^2 \Delta^2 \rangle.
$$
From the relations (\ref{xx})-(\ref{xy}), we obtain the following
expression:
\begin{equation}\label{variance}
\sigma_g^2 = \frac{1}{4} B_x B_y T^2.
\end{equation}
Hence we obtain that the variance of the Dirac phase angle
$$
\sigma_g = \frac{1}{2} \sqrt{B_x B_y} T
$$
grows linearly with the operational time $T$.

\subsection{Ornstein-Uhlenbeck process}

In the discussion concerning the robustness of geometric phases, a
crucial role is played by a typical time scale characterizing the
noise. A time scale is as well needed in order to deal with the
adiabatic limit. For these reasons we consider a model of {\it
colored} noise which presents a typical time scale, namely the
Ornstein-Uhlenbeck stochastic process. Within this model, the
coordinates of noise component fulfill the following stochastic
differential equations:
\begin{eqnarray*}
dx(t) & = & - \Gamma_x x(t) dt + \sqrt{D_x} dW_x(t), \\
dy(t) & = & - \Gamma_y y(t) dt + \sqrt{D_y} dW_y(t),
\end{eqnarray*}
where $W_x(t)$ and $W_y(t)$ are two independent normalized Wiener
processes satisfying
$$
\langle dW_x(s) dW_x(t) \rangle = \langle dW_y(s) dW_y(t) \rangle =
\delta(s-t),
$$
and
$$
\langle dW_x(s) dW_y(t) \rangle = 0.
$$
The two-times correlation function decays exponentially:
\begin{eqnarray*}
\langle x(s) x(t) \rangle & = & \epsilon_x^2 e^{-\Gamma_x|s-t|}, \\
\langle y(s) y(t) \rangle & = & \epsilon_y^2 e^{-\Gamma_y|s-t|},
\end{eqnarray*}
where $\epsilon_x^2=\frac{D_x}{2\Gamma_x}$,
$\epsilon_y^2=\frac{D_y}{2\Gamma_y}$ and $\langle x(s) y(t) \rangle
= 0$.

The Dirac phase angle can be written as the mean square limit (see
e.g.\ \cite{Gardiner}) of the following quantity
$$
\phi_g = \lim_{\mathcal{N}\rightarrow\infty} \mathcal{S} +
\frac{1}{2}\Sigma\Delta,
$$
where
$$
\mathcal{S} = \sum_{j=0}^{\mathcal{N}-1} y(t_j) ( x(t_j + \delta t)
- x(t_j) ),
$$
with $\delta t = T/\mathcal{N}$, and the proper term has been added
to ensure gauge invariance.

The the variance of the Dirac phase angle is given by
\begin{equation} \label{mslimit}
\sigma_g^2 = \lim ( \mathcal{S} + \frac{1}{2}\Sigma\Delta )^2 = \lim
\langle \mathcal{S}^2 \rangle + \lim \langle \mathcal{S} \Sigma
\Delta \rangle + \frac{1}{4} \langle \Sigma^2 \rangle \langle
\Delta^2 \rangle.
\end{equation}
The first term on the right hand side of equation (\ref{mslimit}) is
the limit of
$$
\mathcal{S}^2 = \sum_i \sum_j \langle y(t_i) y(t_j) \rangle \langle
( x(t_i + \delta t) - x(t_i) )( x(t_j + \delta t) - x(t_j) )
\rangle,
$$
where the average of $y$ and $x$ factorizes for the statistical
independence of the processes. We have
$$
\Delta \mathcal{S}^2 = \sum_{ij}\langle y(t_i) y(t_j) \rangle
\left[\langle x(t_{i+1})x(t_{j+1})\rangle - \langle x(t_{i+1}
t)x(t_j)\rangle + \langle x(t_i)x(t_j)\rangle - \langle
x(t_{j+1})x(t_j)\rangle \right]
$$
Evaluating the two-times correlation functions, and putting
$D_x=D_y$ and $\Gamma_x=\Gamma_y$ ,one obtains:
$$
\Delta \mathcal{S}^2 = \epsilon^2 \sum_{ij}\langle y(t_i) y(t_j)
\rangle \left\{ 2e^{-\Gamma |i-j|\delta t} - e^{-\Gamma
|i-j+1|\delta t} - e^{-\Gamma |j-i+1|\delta t} \right\} \;.
$$
The term in curled brackets is
\begin{eqnarray*}
\left\{ \begin{array}{lcc}
2(1-e^{-\Gamma\delta t}) \simeq 2 \Gamma \delta t                         & \mbox{for} & |i-j|=0 \\
-(1 - e^{-\Gamma\delta t})^2 \simeq - \Gamma^2 (\delta t)^2               & \mbox{for} & |i-j|=1 \\
e^{-\Gamma |i-j|\delta t} (2 - e^{\Gamma\delta t} - e^{-\Gamma\delta
t}) \simeq e^{-\Gamma |i-j|\delta t} \Gamma^2 (\delta t)^2 &
\mbox{for} & |i-j| > 1
\end{array}\right. ,
\end{eqnarray*}
and $\langle y(t_i) y(t_j) \rangle = \epsilon^2 e^{-\Gamma
|i-j|\delta t}$. Taking the limit $\delta t \rightarrow 0$, only the
terms with $|i-j|=0$ do not vanish, leading to
$$
\sigma_g^2 \simeq \epsilon^4 \sum_j 2 \Gamma \delta t \simeq
\epsilon^4 \int_0^T 2 \Gamma dt = 2 \epsilon^4 \Gamma T = 2
\epsilon^4 N,
$$
where $N := \Gamma T$ is interpreted as the {\it average number of
statistically independent fluctuations}.

The second term on the right hand side of (\ref{mslimit}) is the
limit of the quantity
$$
\langle  (x(0)-x(T))(y(0)+y(T)) \sum_j y(t_j) (x(t_j+\delta t) -
x(t_j)) \rangle,
$$
which equals
$$
\epsilon^4 \sum_j \left[e^{-\Gamma t_j}+e^{-\Gamma(T-t_j)}\right]
\left[e^{-\Gamma(t_j+\delta t)}-e^{-\Gamma(T-t_j-\delta
t)}-e^{-\Gamma t_j}+e^{-\Gamma(T-t_j)}\right]
$$
and, in the limit of $\delta t \rightarrow 0$ reads:
\begin{eqnarray*}
& & - \Gamma \epsilon^4 \sum_j (e^{-\Gamma t_j}+e^{-\Gamma T}
e^{\Gamma t_j}) (e^{-\Gamma t_j} + e^{-\Gamma T} e^{\Gamma t_j})
\delta t \\
& \simeq & - 4 \Gamma \epsilon^4 e^{-\Gamma T} \int_0^T
\cosh{\left[\Gamma\left(\frac{T}{2}-t\right)\right]}
\cosh{\left[\Gamma\left(\frac{T}{2}-t\right)\right]} dt.
\end{eqnarray*}
The last integral reads
$$
- 2 \epsilon^4 \Gamma \left( \frac{1-e^{-2\Gamma T}}{2\Gamma} + T
\right).
$$

Finally, the third term on the right hand side of (\ref{mslimit}) is
$$
%\frac{1}{4}\left( \langle x(0)^2 \rangle + \langle x(T)^2 \rangle - 2 \langle x(0) x(T) \rangle \right) \left( \langle y(0)^2 \rangle + \langle y(T)^2 \rangle - 2 \langle y(0) y(T) \rangle \right) =
\epsilon^4 \left( 1 - e^{-\Gamma T} \right)^2.
$$

Summing all the contribution, and taking the limit $e^{-\Gamma
T}\rightarrow 0$, we can write:
$$
\sigma^2 \simeq 2 \epsilon^4 \left( \Gamma T + 1 \right).
$$
From the last equation we see that, in contrast to the case of the
Wiener process, for the Orstein-Uhlenbech process the variance
$$
\sigma \simeq \epsilon^2 \sqrt{2 \left( \Gamma T + 1 \right)}.
$$
grows with the square root of the operational time.

\section{Particle in the field of a magnetic monopole}

In this section we consider another simple physical example, in
which the semiclassical particle is subjected to the field of a
magnetic monopole. If the magnetic monopole is sitting at the origin
of the reference frame, we can write the corresponding vector
potential, for $z/R \neq -1$, as follows:
\begin{equation}\label{monopole}
\mathbf{A}\cdot d\mathbf{r} = \frac{1-z/R}{x^2+y^2}\left( -y dx + x
dy \right),
\end{equation}
where $R^2 = x^2 + y^2 + z^2$, or simply in spherical coordinates,
for $\vartheta \neq \pi$, as:
$$
\mathbf{A}\cdot d\mathbf{r} = \frac{1}{R}\tan{\frac{\vartheta}{2}}
d\varphi,
$$
where $\cos{\vartheta}=z/R$ and $\tan{\varphi}=y/x$. For
sufficiently small amplitude of the noise and short operational time
we can take a linearized version of (\ref{monopole}):
\begin{eqnarray*}
\mathbf{A}\cdot d\mathbf{r} \simeq \mathbf{A}_L\cdot d\mathbf{r} & =
& - \left[ f_0 + f_x (x-x_0) + f_y (y-y_0) + f_z (z-z_0) \right] dx \\
& + & \left[ g_0 + g_x (x-x_0) + g_y (y-y_0) + g_z (z-z_0)
\right] dy.
\end{eqnarray*}
In this approximation, we can write the following expression for the
gauge invariant Dirac phase angle:
\begin{eqnarray}
\phi_g & = & - \int \mathbf{A}_L\cdot d\mathbf{r} \nonumber\\
& + & \left[ f_0 - \frac{1}{2} ( f_x \Delta_x + f_y \Delta_y + f_z
\Delta_z ) \right] \Delta_x - \left[ g_0 - \frac{1}{2} ( g_x
\Delta_x + g_y \Delta_y + g_z \Delta_z ) \right] \Delta_y,
\label{mono-dirac}
\end{eqnarray}
where $\Delta_x = x(0) - x(T)$, $\Delta_y = y(0) - y(T)$, $\Delta_z
= z(0) - z(T)$.

\subsection{Wiener process}

In this section we consider a tri-dimensional Wiener process as
noise model. We consider a trivial noiseless loop to which the
Wiener process is superimposed. Denoting as $B_x$, $B_y$ and $B_z$
the diffusion constants, we obtain the following expression for the
mean value of (\ref{mono-dirac}):
$$
\langle \phi_g \rangle = \frac{f_x}{2} \langle \Delta_x^2 \rangle -
\frac{g_y}{2} \langle \Delta_y^2 \rangle = \frac{1}{2}(g_y B_y - f_x
B_x) T,
$$
which does not vanish in general and grows linearly with the
operational time.

Regarding the corresponding variance, we obtain:
$$
\sigma_g^2 = \langle \phi_g^2 \rangle -  \langle \phi_g \rangle^ 2 =
\frac{1}{4} \mathcal{B} T^2.
$$
Hence the variance
$$
\sigma_g = \frac{1}{2}\sqrt{\mathcal{B}} T,
$$
grows linearly with the operational time, and
$$
\mathcal{B} = B_x(3 B_x f_x^2 + 2 B_y f_y^2 + 2 B_z f_z^2) + B_y(2
B_x g_x^2 + 3 B_y g_y^2 +  2 B_z g_z^2).
$$

\subsection{Ornstein-Uhlenbeck process}

In the examples discussed above we computed the variance of the
Dirac phase angle caused by a Brownian motion of the semiclassical
particle. We explicitly considered the case of a trivial noiseless
loop, in which the trajectory is purely Brownian. We obtained that
the leading term in the variance is of the second order in the
amplitude of the noise. Moreover, it increases linearly with the
operational time for the Wiener process, and with the square root in
the case of the Ornstein-Uhlenbeck process. This behavior can be
compared with the results presented in \cite{DeChiara}, where the
variance of the Berry phase was computed in the presence of a noise
component modeled by an Ornstein-Uhlenbeck process. This noisy
component is superimposed to a drift loop which is a precession
around the $z$ axis. In that case it was shown that the leading term
in the variance is of the first order in the amplitude of the noise.
Moreover, it decreases linearly with the operational time, leading
to negligible fluctuations in the Berry phase.

\begin{figure}[htbp]
\centering
\includegraphics[width=0.8\textwidth]{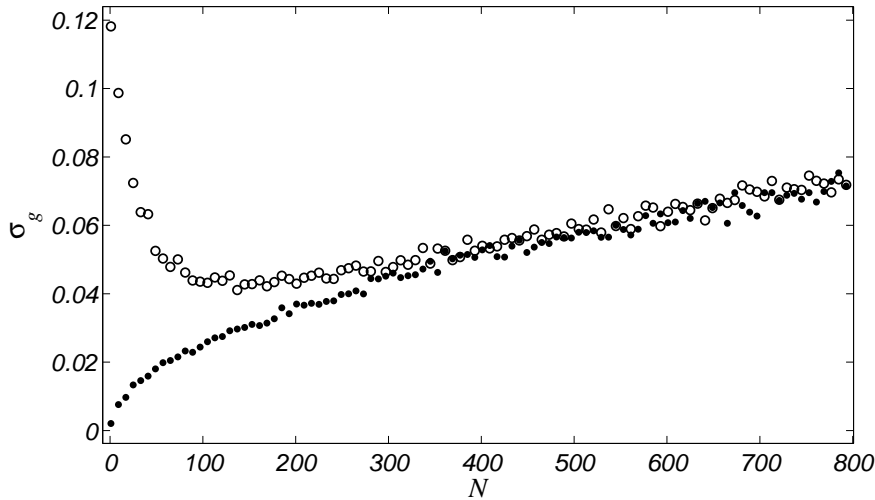}
\caption{The plot shows the numerically estimated variance of the
geometric phase angle for a particle in a monopole field. The
variance is plotted as a function of the average number of noise
fluctuation for a Ornstein-Uhlenbeck process modeling the noise
component, with amplitude $\epsilon=0.05$. The data represented by
dots refer to a trivial drift loop (purely Brownian motion), they
are well fitted by a square root law $\sigma_g = a \sqrt{N} + b$,
with $a\simeq 0.0025$ and $b=-0.00016$. The data represented by
circles correspond to the drift loop in equation (\ref{drift})
(Brownian motion superimposed to precession).} \label{data3}
\end{figure}

It is interesting to compare the variances of the Dirac phase
obtained in the case of different drift loops. We have considered
both the case of a trivial noiseless component
$$
\mathbf{r}_0(t)=(\sin{(\vartheta_0)}\cos{(\varphi_0)},\sin{(\vartheta_0)}\sin{(\varphi_0)},\cos{(\vartheta_0)}),
$$
and the case of a precession about the $z$ axis described by the
loop
\begin{equation}\label{drift}
\mathbf{r}_0(t) = (\sin{(\vartheta_0)}\cos{(\varphi_0+2\pi
t)},\sin{(\vartheta_0)}\sin{(\varphi_0+2\pi
t)},\cos{(\vartheta_0)}).
\end{equation}
We have numerically simulated (following \cite{update}) an
Ornstein-Uhlenbeck process affecting these loops and estimated the
variance of the corresponding Dirac phase angle. The results are
plotted in figure \ref{data3} for $\cos\vartheta_0=1/\sqrt{3}$.

We notice two different pattern of the variance of the Dirac phase
angle as function of the average number of fluctuations in the noisy
component. In the case of trivial drift loop (purely Brownian
motion) the variance always increases as the square root of the
average number of fluctuations. On the other hand, for nontrivial
noiseless loop (Brownian component superimposed to precession) a
transient behavior is present in which the variance decreases with
the number of fluctuations. This behavior is in agreement to what
was found in \cite{DeChiara} and is due to contribution of the first
order in the noise amplitude. By increasing the value of $N$, the
first order contributions become negligible while the second order
ones become predominant.

\section{Conclusion}

We have computed the mean value and the variance of the Dirac phase
acquired by a semiclassical particle subjected to Brownian motion.
If the trajectory is purely Brownian the variance of the Dirac phase
angle always increases as function of the operational time (or the
average number of noise fluctuations). On the other hand, a
transient behavior is observed if the Brownian motion is
superimposed to a noiseless drift loop.

In the case of pure Brownian motion, we have obtained an expression
for the variance which is of the second order in the amplitude of
noise and increases with the operational time. In particular, if the
noise is modeled by a Ornstein-Uhlenbeck process the variance grows
with the square root of the operational time.

The case of the Dirac phase can be viewed as instance of geometric
phase. Hence we can compare our results to others which refer to
Berry phase. In \cite{DeChiara}, it was shown that in the case of an
adiabatic precession of a $1/2$-spin the leading term in the
variance of the Berry phase is of the first order in the amplitude
of the noise. Moreover, this terms decrease linearly with the
operational time. This behavior is in accordance with the transient
behavior of the Dirac phase for nontrivial noiseless drift loop.
Notice that the second order effects become relevant for long enough
operational time.

At the best of our knowledge, the presentation of the effects of
second order in the variance of the Dirac phase introduces a new
element in the study of the robustness of geometric phases. We argue
that second order effects are feasible to be observed in
experimental settings as in \cite{Filipp}. Recently, the effects of
non-adiabaticity in the noise component were studied in \cite{Hou}.
The pattern of the variance of the corresponding geometric phase as
function of the operational time is qualitative analogous to the
results presented here, in the sense that the squared variance grows
linearly in time. Quantitatively, this effect is of the first order
in the noise amplitude in \cite{Hou} while in our analysis, which
assumes the adiabatic approximation, the effect of the second order.


\begin{thebibliography}{99}

\bibitem{Panch} S. Pancharatnam,
%{\it Generalized theory of interference, and its applications}
{\it Proc. Indian Acad. Sci. A} {\bf 44} 247 (1956); reprinted in
{\it Collected works of S. Pancharatnam} (Oxford Univ. Press,
London, 1975)

\bibitem{Berry} M. V. Berry,
%{\it Quantal Phase Factors Accompanying Adiabatic Changes}
{\it Proc. R. Soc. Lond. A} {\bf 392} 45 (1984)

\bibitem{Simon} B. Simon,
%{\it Holonomy, the Quantum adiabatic Theorem, and Berry's Phase}
{\it Phis. Rev. Lett.} {\bf 51} 2167 (1983)

\bibitem{Wilczek} F. Wilczek and A. Zee,
%{\it Appearance of Gauge Structure in Simple Dynamical Systems}
{\it Phis. Rev. Lett.} {\bf 52} 2111 (1984)

\bibitem{Aharonov} Y. Aharonov, J. Anandan,
%{\it Phase change during a cyclic quantum evolution}
{\it Phis. Rev. Lett.} {\bf 58} 1593 (1987); \\
%
J. Anandan,
%{\it Non-adiabatic non-abelian geometric phase}
{\it Phys. Lett. A} {\bf 133} 171 (1988)

\bibitem{Samuel}
J. Samuel and R. Bhandari,
%{\it General Setting for Berry's Phase}
{\it Phis. Rev. Lett.} {\bf 60} 2339 (1988)

\bibitem{Bohm} A. Bohm, A. Mostafazadeh, H. Koizumi, Q. Niu and
J. Zwanziger \textit{The Geometric Phase in Quantum Systems}
(Heidelberg: Springer-Verlag, 2003)

\bibitem{Bennett} C. H. Bennett and G. Brassard,
{\it Quantum Cryptography: Public Key Distribution and Coin
Tossing}, Proceedings of IEEE International Conference on Computers
Systems and Signal Processing, Bangalore India (1984); \\
%
A. K. Ekert,
%{\it Quantum cryptography based on Bell's theorem}
{\it Phis. Rev. Lett.} {\bf 67} 661 (1991); \\
%
P. W. Shor,
%{\it Polynomial-Time Algorithms for Prime Factorization and Discrete Logarithms on a Quantum Computer}
{\it SIAM J. Sci. Statist. Comput.} {\bf 26} 1484 (1997)

\bibitem{Nielsen_Chuang} M. A. Nielsen, I. L. Chuang,
\textit{Quantum Computation and Quantum Information} (Cambridge
University Press, Cambridge, 2000)

\bibitem{Zanardi_Rasetti} P. Zanardi, M. Rasetti,
%{\it Holonomic Quantum Computation}
{\it Phys. Lett. A} {\bf 264} 94 (1999)

\bibitem{Deutsch} D. Deutsch,
%{\it Quantum theory, the Churh-Turing principle and the universal quantum computer}
{\it Proc. R. Soc. Lond. A} {\bf 400} 97 (1985)

\bibitem{Falci} G. Falci, R. Fazio, G. M. Palma, J. Siewert,
and V. Vedral,
%{\it Detection of geometric phases in superconducting nanocircuits}
{\it Nature} {\bf 407} 355 (2000)

\bibitem{Duan} L. M. Duan, J. I. Cirac, P. Zoller,
%{\it Geometric Manipulation of Trapped Ions for Quantum Computation}
{\it Science} {\bf 292} 1695 (2001)

\bibitem{nist} D. Leibfried, B. De Marco, V. Meyer, D. Lucas, M. Barrett, J. Britton, W. M. Itano, B. Jelenkovic, C. Langer, T. Rosenband, and
D. J. Wineland,
%{\it Experimental demonstration of a robust, high-fidelity geometric two ion-qubit phase gate}
{\it Nature} {\bf 422} 412 (2003)

\bibitem{montana} M. Tian, Z. W. Barber, J. A. Fischer, and W. Randall
Babbitt,
%{\it Geometric manipulation of the quantum states of two-level atoms}
{\it Phys. Rev. A} {\bf 69} 050301(R) (2004)

\bibitem{Jones} J. A. Jones, V. Vedral, A. Ekert and G. Castagnoli,
%{\it Geometric quantum computation using nuclear magnetic resonance}
{\it Nature} {\bf 403} 869 (2000)

\bibitem{ECC} P. W. Shor,
%{\it Scheme for reducing decoherence in quantum computer memory}
{\it Phys. Rev. A} {\bf 52} 2493 (1995); \\
%
A. Steane,
%{\it Multiple Particle Interference and Quantum Error Correction}
{\it Proc. R. Soc. Lond. A} {\bf 452} 2551 (1996); \\
%
A. R. Calderbank and P. W. Shor,
%{\it Good quantum error-correcting codes exist}
{\it Phys. Rev. A} {\bf 54} 1098 (1996)

\bibitem{Kitaev} A. Y. Kitaev,
%{\it Fault-tolerant quantum computation by anyons}
{\it Ann. Phys.} {\bf 303} 2 (2003)

\bibitem{DeChiara} G. De Chiara, G. M. Palma,
%{\it Berry Phase for a Spin $1/2$ Particle in a Classical Fluctuating Field}
{\it Phys. Rev. Lett.} {\bf 91} 090404 (2003)

\bibitem{Solinas} P. Solinas, P. Zanardi and N. Zangh\`i,
%{\it Robustness of non-Abelian holonomic quantum gates against parametric noise}
{\it Phys. Rev. A} {\bf 70} 042316 (2004); \\
%
G. Florio, P. Facchi, R. Fazio, V. Giovannetti and S. Pascazio,
%{\it Robust gates for holonomic quantum computation}
{\it Phys. Rev. A} {\bf 73} 022327 (2006); \\
%
C. Lupo, P. Aniello, M. Napolitano, G. Florio,
%{\it  Robustness against parametric noise of nonideal holonomic gates}
{\it Phs. Rev. A} 76 012309 (2007)

\bibitem{Filipp} S. Filipp, J. Klepp, Y. Hasegawa, C. Plonka-Spehr, U. Schmidt, P. Geltenbort,
H. Rauch, arXiv:0812.3757v1 (2008)

\bibitem{Gardiner} C. W. Gardiner,
{\it Handbook of Stochastic Methods} (Springer, Berlin, 1983)

\bibitem{update} D. T. Gillespie,
%
{\it Phys. Rev. E} {\bf 54} 2084 (1996)

\bibitem{Hou} X. J. Hou,
%{\it Effect of classical noise on the geometric quantum phase}
{\it Phys. Rev. A} {\bf 75} 024103 (2007)

\end{thebibliography}
\end{document}